\documentclass[showpacs,reprint]{revtex4}
\usepackage{amsfonts}
\usepackage{amssymb}
\usepackage{amsmath}
\usepackage{hyperref}
\usepackage{latexsym}
\usepackage[dvips]{graphicx}
\usepackage{epsf}
\usepackage{multirow}

\begin{document}

\title{Gas of wormholes in Euclidean quantum field theory}
\author{ E.P. Savelova}
\email{sep$\_$ 22.12.79@inbox.ru}
\affiliation{Dubna International University of Nature, Society and Man, Universitetskaya
Str. 19, Dubna, 141980, Russia }
\date{}

\begin{abstract}
We model the spacetime foam picture by a gas of wormholes in Euclidean field
theory. It is shown that at large distances the presence of such a gas leads
merely to a renormalization of mass and charge values. We also demonstrate
that there exist a class of specific distributions of point-like wormholes
which essentially change the ultraviolet behavior of Green functions and
lead to finite quantum field theories.
\end{abstract}

\maketitle

\section{Introduction}

It is widely expected that at very small scales spacetime has a foam-like
structure \cite{H78} and from the very beginning it was expected that the
spacetime foam picture may solve the problem of divergencies in quantum
field theory. The most rigorous description of the spacetime foam can be
achieved in the approach which uses the lattice quantum gravity e.g., see
\cite{AJL,MRS} and references therein. In particular, it allows to calculate
the spectral dimension for nonperturbative quantum gravity \cite{LQG} which
seems to run from a value of $D=3/2$ at short distance to $D=4$ at large
distance scales. However, such an approach requires numerical investigations
which are not sufficiently visual. One would like to have a picture in which
basic properties of the spacetime foam become more transparent and may allow
to control numerical investigations.

We point out that the spacetime foam structure represents quantum topology
fluctuations and roughly it can be described by the picture when at very
small scales space is filled with a gas of virtual wormholes. Unlike actual
wormholes virtual wormhole exists only for a very short period of time and
at very small scales, and does not obey to the Einstein equations and,
therefore, there are no theorems which may forbid origin of such objects.
From the mathematical standpoint the simplest virtual wormhole can be
described by a standard four-dimensional wormhole whose throat section is a
3-dimensional sphere $S^{3}$ and due to its high symmetry such an object
admits the rigorous mathematical treatment (analogous to \cite{KS07}). In
the present paper we model the spacetime foam by a gas of wormholes in
Euclidean field theory and investigate the simplest spacetime foam effects.

We point out that the interest to Euclidean wormholes is not new. There was
a wide investigation of Euclidean wormholes or baby universes in connection
to topology changes and the loss of quantum coherence \cite{H88,lrt88}, see
also \cite{loss}. The results of observations of neutrino oscillations and
other oscillations \footnote{%
Loss of coherence must lead to damping of oscillations} ($\nu _{\tau
}\leftrightarrow \nu _{\mu }$, $K\leftrightarrow K$, etc.) \cite{exp}
suggested that there are severe constraints on the magnitude of processes of
the actual changing the topology as it was predicted by \cite{lrt88} and
that such processes at the present time are strongly suppressed. Those
however do not forbid the origin of wormhole-like structures as discussed by
\cite{KS12}. In turn, this also means that changes in the topology in the
modern Universe can be only virtual (i.e., the space-time foam).

\section{The structure of a wormhole}

In the Euclidean field theory the simplest wormhole can be constructed
merely by a gluing procedure \cite{KS10} and is described by the metric ($%
\alpha =1,2,3,4$)%
\begin{equation}
ds^{2}=h^{2}(r)\delta _{\alpha \beta }dx^{\alpha }dx^{\beta },  \label{wmetr}
\end{equation}%
where $h(r)=1+\theta \left( a-r\right) \left( \frac{a^{2}}{r^{2}}-1\right) $
and $\theta \left( x\right) $ is the step function. Such a wormhole has
vanishing throat length. Indeed, in the region $r>a$, $h=1$ and the metric
is flat, while the region $r<a$, with the obvious transformation $y^{\alpha
}=\frac{a^{2}}{r^{2}}x^{\alpha }$, is also flat for $y>a$. Therefore, the
regions $r>a$ and $r<a$ represent two Euclidean spaces glued at the surface
of a sphere $S^{3}$ with the centre at the origin $r=0$ and the radius $r=a$%
. Such a space can be described with the ordinary double-valued flat metric
in the region $r_{\pm }>a$ by
\begin{equation}
ds^{2}=\delta _{\alpha \beta }dx_{\pm }^{\alpha }dx_{\pm }^{\beta },
\label{wmetr2}
\end{equation}%
where the coordinates $x_{\pm }^{\alpha }$ describe two different sheets of
space. Now, identifying the inner and outer regions of the sphere $S^{3}$
allows the construction of a wormhole which connects regions in the same
space (instead of two independent spaces). This is achieved by gluing the
two spaces in (\ref{wmetr2}) by motions of the Euclidean space. If $R_{\pm }$
is the position of the sphere in coordinates $x_{\pm }^{\mu }$, then the
gluing is the rule%
\begin{equation}
x_{+}^{\mu }=R_{+}^{\mu }+\Lambda _{\nu }^{\mu }\left( x_{-}^{\nu
}-R_{-}^{\nu }\right) ,  \label{gl}
\end{equation}%
where $\Lambda _{\nu }^{\mu }\in O(4)$, which represents the composition of
a translation and a rotation of the Euclidean space (Lorentz
transformation). In terms of common coordinates such a wormhole represents
the standard flat space in which the two spheres $S_{\pm }^{3}$ (with
centers at positions $R_{\pm }$) are glued by the rule (\ref{gl}). We point
out that the physical region is the outer region of the two spheres. Thus,
in general, the wormhole is described by a set of parameters: the throat
radius $a$, positions of throats $R_{\pm }$, and the rotation matrix $%
\Lambda _{\nu }^{\mu }\in O(4)$.

\section{The Green function for a single wormhole connecting two spaces}

In the Euclidean field theory the Green function obeys the Laplace equation
\begin{equation}
\left( -\Delta +\zeta R+m^{2}\right) G(x,x^{\prime })=4\pi ^{2}\delta
(x-x^{\prime })  \label{weq}
\end{equation}%
where $m$ is the mass, $R$ is the scalar curvature, $\Delta =\frac{1}{\sqrt{g%
}}\partial _{\alpha }(\sqrt{g}g^{\alpha \beta }\partial _{\beta })$, and $%
g_{\alpha \beta }$ is the metric. In the Euclidean space the metric is flat $%
g_{\alpha \beta }=\delta _{\alpha \beta }$ and the above equation has the
well-known solution $G_{0}(x,x^{\prime })=m^{2}\frac{K_{1}(mr)}{mr}$, where $%
r^{2}=(x-x^{\prime })^{2}$ and $K_{1}(x)$ is the modified Bessel function.
When considering a space with a wormhole the metric cannot be chosen
everywhere flat and the exact form of the Green function depends on the
specific structure of the wormhole. In the present section we construct the
Green function for the simplest wormhole connecting two Euclidean spaces
which is described by the metric (\ref{wmetr}). In this case the equation (%
\ref{weq}) admits the exact solution.

Indeed, consider four-dimensional spherical coordinates $r$, $\chi $, $%
\theta $, $\phi $ connected to the cartesian coordinates by

\begin{eqnarray}
x &=&r\sin \chi \sin \theta \cos \phi ,~y=r\sin \chi \sin \theta \sin \phi
\label{sp} \\
z &=&r\sin \chi \cos \theta ,\ t=r\cos \chi .  \notag
\end{eqnarray}%
where $0\leq r<\infty $, $0\leq \phi <2\pi $, and $0\leq \chi $, $\theta
<\pi $. The square of the element of length in this coordinates is%
\begin{equation*}
ds^{2}=h^{2}\left( r\right) \left( dr^{2}+r^{2}\left[ d\chi ^{2}+\sin
^{2}\chi \left( d\theta ^{2}+\sin ^{2}\theta d\phi ^{2}\right) \right]
\right) .
\end{equation*}%
For the Laplacian operator in spherical coordinates in a four-dimensional
Euclidean space applied to a scalar $f$ we get
\begin{equation*}
\Delta f=\frac{1}{r^{3}h^{4}}\frac{\partial }{\partial r}\left( r^{3}h^{2}%
\frac{\partial f}{\partial r}\right) +\frac{1}{r^{2}h^{2}}\Delta _{\Omega }f
\end{equation*}%
where $\Delta _{\Omega }f=f_{;\alpha }^{;\alpha }$ is the angular part of $%
\Delta $ ($\Omega =(\chi ,\theta ,\phi )$). The angular part of $f$ can be
written in terms of four -dimensional spherical harmonics $f=\sum
f_{nlm}Q_{nlm}$ e.g., see Ref. \cite{fock} where%
\begin{equation*}
\Delta _{\Omega }Q_{nlm}=-(n^{2}-1)Q_{nlm},
\end{equation*}%
($n=1,2,...$) which can be decomposed as $Q_{nlm}(\Omega
)=q_{lm}^{(n)}Y_{lm}(\theta ,\phi )\Pi _{nl}(\chi )$, where $q_{lm}^{(n)}$
are normalization constants, $Y_{lm}(\theta ,\phi )$ are the usual
three-dimensional spherical harmonics, and
\begin{equation*}
\Pi _{nl}(\chi )=\sin ^{l}\chi \frac{d^{l+1}(\cos n\chi )}{d(\cos \chi
)^{l+1}},~(l=0,1,...,n-1).
\end{equation*}%
(In particular, the most symmetric scalar spherical harmonic corresponds to $%
l=0$ and has the form $Q=\frac{\sin n\chi }{\sin \chi }$, $n=1,2,3,...$).
These functions obey the normalization conditions%
\begin{equation*}
\int_{0}^{\pi }\sin ^{2}\chi d\chi \int_{0}^{\pi }\sin \theta d\theta
\int_{0}^{2\pi }d\phi Q_{nlm}^{\ast }(\Omega )Q_{n^{\prime }l^{\prime
}m^{\prime }}(\Omega )=\delta _{nn^{\prime }}\delta _{ll^{\prime }}\delta
_{mm^{\prime }}
\end{equation*}%
and the completeness condition
\begin{equation*}
\sum_{n=1}^{\infty }\sum_{l=0}^{n-1}\sum_{m=-l}^{l}Q_{nlm}^{\ast }(\Omega
)Q_{nlm}(\Omega ^{\prime })=\frac{1}{\sin ^{2}\chi \sin \theta }\delta
(\Omega -\Omega ^{\prime }).
\end{equation*}%
The Euclidean Green function (for $m=0$ case) has the decomposition%
\begin{equation*}
G_{0}(x,x^{\prime })=\frac{1}{(x-x^{\prime })^{2}}=\sum_{n=1}^{\infty }\frac{%
x_{<}^{n-1}}{x_{>}^{n+1}}Q_{n}(\widetilde{\chi }).
\end{equation*}%
where $x_{>}$ and $x_{<}$ are the biggest and smallest value of $x$, $%
x^{\prime }$ respectively and we have used the composition theorem for
spherical harmonics
\begin{equation}
Q_{n}(\widetilde{\chi })=\frac{4\pi ^{2}}{2n}\sum_{l=0}^{n-1}%
\sum_{m=-l}^{l}Q_{nlm}^{\ast }(\Omega ^{\prime })Q_{nlm}(\Omega )
\label{com}
\end{equation}%
where $\widetilde{\chi }$ is the angle between the vectors $x$ and $%
x^{\prime }$.

Consider the decomposition of $\delta $-function as%
\begin{equation*}
\delta (x-x^{\prime })=\frac{1}{r^{3}}\delta (r-r^{\prime
})\sum_{n=1}^{\infty }\sum_{l=0}^{n-1}\sum_{m=-l}^{l}Q_{nlm}^{\ast }(\Omega
)Q_{nlm}(\Omega ^{\prime }).
\end{equation*}%
Then we may represent the Green function in the same form%
\begin{equation*}
G(x,x^{\prime })=\sum_{n=1}^{\infty
}\sum_{l=0}^{n-1}\sum_{m=-l}^{l}Q_{nlm}^{\ast }(\Omega ^{\prime
})g_{n}(r,r^{\prime })Q_{nlm}(\Omega )
\end{equation*}%
which gives from (\ref{weq})%
\begin{equation}
-\frac{1}{r^{3}h^{4}}\frac{\partial }{\partial r}\left( r^{3}h^{2}\frac{%
\partial g_{n}}{\partial r}\right) +(\zeta R+m^{2})g_{n}+\frac{(n^{2}-1)}{%
r^{2}h^{2}}g_{n}=\frac{4\pi ^{2}}{r^{3}h^{4}}\delta (r-r^{\prime }).
\label{weq2}
\end{equation}%
Here the scalar curvature with the metric (\ref{wmetr}) is \cite{KS10} (the
step function at the junction requires considering distributional curvature
etc., see details in \cite{taub})
\begin{equation*}
R=-\frac{12}{a}\delta \left( r-a\right) .
\end{equation*}

Consider the region $r>a$ which corresponds to the first sheet of the
Euclidean space. Then $h=1$ and if $r\neq r^{\prime }$ the above equation
gives%
\begin{equation*}
\frac{\partial ^{2}g_{n}}{\partial x^{2}}+\frac{3}{x}\frac{\partial g_{n}}{%
\partial x}-\left( 1+\frac{(n^{2}-1)}{x^{2}}\right) g_{n}=0
\end{equation*}%
where $x=mr$. The substitution $g=f/x$ transforms it into the modified
Bessel equation
\begin{equation*}
\frac{\partial ^{2}f}{\partial x^{2}}+\frac{1}{x}\frac{\partial f}{\partial x%
}-\left( 1+\frac{n^{2}}{x^{2}}\right) f=0
\end{equation*}%
whose solutions are the modified Bessel functions $I_{n}(x)$ and $K_{n}(x)$
and therefore we find%
\begin{equation}
g_{n}=C_{1}\frac{I_{n}(mr)}{mr}+C_{2}\frac{K_{n}(mr)}{mr}.  \label{g}
\end{equation}%
In the region $r<a$ we use the coordinate transformation $y^{\alpha }=\frac{%
a^{2}}{r^{2}}x^{\alpha }$ which does not change the angular dependence,
while it transforms $r\rightarrow \widetilde{r}=a^{2}/r$. Thus the region $%
r<a$ corresponds to the second sheet of the Euclidean space $\widetilde{r}>a$
where we get the same solution (\ref{g}), or%
\begin{equation}
g_{n}=\widetilde{C}_{1}\frac{I_{n}(ma^{2}/r)}{ma^{2}/r}+\widetilde{C}_{2}%
\frac{K_{n}(ma^{2}/r)}{ma^{2}/r}.  \label{g2}
\end{equation}

We recall that these functions has the well-known asymptotic as $%
x\rightarrow 0$
\begin{equation*}
\frac{I_{n}(x)}{x}\rightarrow \frac{1}{2\Gamma \left( n+1\right) }\left(
\frac{x}{2}\right) ^{n-1},\frac{~K_{n}(x)}{x}\rightarrow \frac{\Gamma \left(
n\right) }{4}\left( \frac{2}{x}\right) ^{n+1}
\end{equation*}%
and as $x\rightarrow \infty $%
\begin{equation*}
I_{n}(x)\rightarrow e^{x}\left( \frac{1}{\sqrt{2\pi x}}+o(x^{-3/2})\right)
,~K_{n}(x)\rightarrow e^{-x}\sqrt{\frac{\pi }{2x}}\left( 1+o(\frac{1}{x}%
)\right) .
\end{equation*}

Now to get the solution to (\ref{weq2}) \ we have to consider the three
regions $0<r<a$, \ $a<r<r^{\prime }$, and $r>r^{\prime }$ where the solution
has the form (\ref{g2}) and (\ref{g}) respectively, and match this functions
by the boundary conditions. At the boundary $r=a$ and at the source $%
r=r^{\prime }$we get
\begin{equation*}
\frac{\partial g_{n}}{\partial r}\Big|_{a-0}^{a+0}=-\zeta \frac{12}{a}%
h^{2}g_{n}(a)
\end{equation*}%
\begin{equation*}
\frac{\partial g_{n}}{\partial r}\Big|_{r^{\prime }-0}^{r^{\prime }+0}=-%
\frac{4\pi ^{2}}{r^{\prime }{}^{3}h^{2}}
\end{equation*}%
\begin{equation*}
g_{n}(a-0)=g_{n}(a+0)
\end{equation*}%
\begin{equation*}
g_{n}(r^{\prime }-0)=g_{n}(r^{\prime }+0),
\end{equation*}%
while the two more boundary conditions we get from the requirement $%
g\rightarrow 0$ as $r,\widetilde{r}\rightarrow \infty $, which give as $r<a$
\begin{equation*}
g_{n}=\widetilde{C}\frac{K_{n}\left( m\frac{a^{2}}{r}\right) }{m\frac{a^{2}}{%
r}}
\end{equation*}%
and as $r>r^{\prime }$
\begin{equation*}
g_{n}=C\frac{K_{n}\left( mr^{\prime }\right) }{mr^{\prime }}.
\end{equation*}%
The jump of derivatives at positions $r=r^{\prime }$ and $r=a$ means the
presence of sources. At $r=r^{\prime }$ $\ $we have the standard external
source from the right hand side of (\ref{weq2}), while at $r=a$ the source
generates due to the curvature. The explicit form of the function $g_{n}$ in
every of regions is
\begin{equation*}
g_{n}=\left\{
\begin{array}{ll}
A\frac{K_{n}\left( m\frac{a^{2}}{r}\right) }{m\frac{a^{2}}{r}}\, & \hbox{as}%
\,\,\,0<r<a; \\
B\frac{I_{n}\left( mr\right) }{mr}+C\frac{K_{n}\left( mr\right) }{mr}, & %
\hbox{as}\,\,\,a<r<r^{\prime }; \\
D\frac{K_{n}\left( mr\right) }{mr}, & \hbox{as}\,\,\,r^{\prime }<r<\infty ;%
\end{array}%
\right.
\end{equation*}%
where the coefficients are expressed as
\begin{equation*}
B=4\pi ^{2}m^{2}\frac{K_{n}(mr^{\prime })}{mr^{\prime }}
\end{equation*}%
\begin{equation*}
C=-\frac{\left( \ln \left( \frac{I_{n}(ma)}{ma}\frac{K_{n}(ma)}{ma}\right)
\right) ^{\prime }+\frac{12}{a}\zeta }{2\left( \left( \ln \frac{K_{n}(ma)}{ma%
}\right) ^{\prime }+\frac{6}{a}\zeta \right) }\frac{\frac{I_{n}(ma)}{ma}}{%
\frac{K_{n}(ma)}{ma}}B
\end{equation*}%
\begin{equation*}
D=C+4\pi ^{2}m^{2}\frac{I_{n}(mr^{\prime })}{mr^{\prime }}
\end{equation*}%
\begin{equation*}
A=-C-\frac{\left( \ln \frac{I_{n}(ma)}{ma}\right) ^{\prime }+\frac{6}{a}%
\zeta }{\left( \ln \frac{K_{n}(ma)}{ma}\right) ^{\prime }+\frac{6}{a}\zeta }%
\frac{\frac{I_{n}(ma)}{ma}}{\frac{K_{n}(ma)}{ma}}B
\end{equation*}%
where $\left( \frac{K_{n}(ma)}{ma}\right) ^{\prime }=\frac{\partial }{%
\partial a}\frac{K_{n}(ma)}{ma}$. Thus we find for the Green function the
expressions as $r<a$%
\begin{equation*}
g_{n}=\left( \beta _{n}(a,r^{\prime })-\alpha _{n}(a,r^{\prime })\right)
g_{n}^{0}\left( \widetilde{r},a\right)
\end{equation*}%
and as $r>a$%
\begin{equation*}
g_{n}=g_{n}^{0}\left( r,r^{\prime }\right) +\alpha _{n}(a,r^{\prime
})g_{n}^{0}\left( r,a\right)
\end{equation*}%
where
\begin{equation}
\alpha _{n}(a,r^{\prime })=-\frac{\left( \ln \left( \frac{I_{n}(ma)}{ma}%
\frac{K_{n}(ma)}{ma}\right) \right) ^{\prime }+\frac{12}{a}\zeta }{2\left(
\left( \ln \frac{K_{n}(ma)}{ma}\right) ^{\prime }+\frac{6}{a}\zeta \right) }%
\frac{\frac{K_{n}(mr^{\prime })}{mr^{\prime }}}{\frac{K_{n}(ma)}{ma}},
\label{a}
\end{equation}%
\begin{equation}
\beta _{n}(a,r^{\prime })=-\frac{\left( \ln \frac{I_{n}(ma)}{ma}\right)
^{\prime }+\frac{6}{a}\zeta }{\left( \ln \frac{K_{n}(ma)}{ma}\right)
^{\prime }+\frac{6}{a}\zeta }\frac{\frac{K_{n}(mr^{\prime })}{mr^{\prime }}}{%
\frac{K_{n}(ma)}{ma}},  \label{b}
\end{equation}%
and $g_{n}^{0}$ is the standard Euclidean Green function which is given by
\begin{equation}
g_{n}^{0}\left( r,r^{\prime }\right) =4\pi ^{2}m^{2}\frac{K_{n}(mr_{>})}{%
mr_{>}}\frac{I_{n}(mr_{<})}{mr_{<}}  \label{gf0}
\end{equation}%
where $r_{>}$ and $r_{<}$ denote the biggest and smallest value of $r$ and $%
r^{\prime }$ respectively.

In the massless case $m=0$ we find \cite{KS10} as $r<a$%
\begin{equation*}
g_{n}=\left( \frac{6\zeta -1}{\left( 6\zeta -(n+1)\right) }-\frac{6\zeta +n-1%
}{6\zeta -(n+1)}\right) \left( \frac{a}{r^{\prime }}\right) ^{n+1}g_{n}^{0}(%
\widetilde{r},a)
\end{equation*}%
and as $r>a$%
\begin{equation*}
g_{n}^{0}+\Delta g_{n}=g_{n}^{0}(r,r^{\prime })-\frac{6\zeta -1}{\left(
6\zeta -(n+1)\right) }\left( \frac{a}{r^{\prime }}\right)
^{n+1}g_{n}^{0}(r,a)
\end{equation*}%
where $g_{n}^{0}$ is given by (\ref{gf0}). Or explicit expressions as $r<a$%
\begin{equation*}
g_{n}=\frac{4\pi ^{2}}{2n}\frac{n}{n+1-6\zeta }\left( \frac{a^{2}}{r^{\prime
}\widetilde{r}}\right) ^{n-1}\left( \frac{a}{r^{\prime }\widetilde{r}}%
\right) ^{2}
\end{equation*}%
and as $r>a$%
\begin{equation*}
g_{n}^{0}+\Delta g_{n}=\frac{4\pi ^{2}}{2n}\left( \frac{1}{r_{>}}\right)
^{2}\left( \frac{r_{<}}{r_{>}}\right) ^{n-1}-\frac{4\pi ^{2}}{2n}\frac{%
1-6\zeta }{n+1-6\zeta }\left( \frac{a^{2}}{rr^{\prime }}\right) ^{n-1}\left(
\frac{a}{r^{\prime }r}\right) ^{2}.
\end{equation*}%
\ In particular, conformal coupling corresponds to $\zeta =1/6$ and in the
region $r>a$ we get $\Delta g_{n}=0$ as it should be. In this case the Green
function can be constructed by the image method (e.g., in three dimensions
such an approach was used in \cite{KS07}) or by conformal transformation of
the standard Euclidean Green function. In the case of minimal coupling $%
\zeta =0$\ we get the expressions presented in  \cite{KS10}.

In what follows we shall compare this solution with the case when the
wormhole is absent. Then the total space consists of a couple of sheets of
the standard Euclidean spaces $R_{+}^{4}$ and $R_{-}^{4}$ and the total
Green function consists also of the two parts $G_{0}^{+}$ and $G_{0}^{-}$.
Let the source belong to the first space $x^{\prime }\in R_{+}^{4}$. Then we
get $G_{0}^{-}=0$, while $G_{0}^{+}$ corresponds to the standard Euclidean
Green function $G_{0}$, i.e. (\ref{gf0}). In the presence of the wormhole it
is convenient to extract the part related to the standard Green function,
i.e., to define the difference $G=G_{0}+\Delta G$. Then in the region $r>a$
(i.e., as $x=x_{+}\in R_{+}^{4}$) and in the region $r<a$ (i.e., $\widetilde{%
r}>a$, or $x=x_{-}\in R_{-}^{4}$) we get the expression%
\begin{equation*}
\Delta g_{n}^{\pm }=\frac{1}{2}\left( \beta _{n}(a,r^{\prime })\pm \left(
2\alpha _{n}(a,r^{\prime })-\beta _{n}(a,r^{\prime })\right) \right)
g_{n}^{0}\left( r,a\right) .
\end{equation*}

\section{The bias of the source, partial amplitudes}

In this section we find the charge density generated on the surface of the
wormhole throat. To this end we consider the equivalent problem. Indeed the
true Green function obeys the equation (we recall that in the Euclidean
space $R=0$)%
\begin{equation*}
-(\Delta -m^{2})G_{+}\left( x,x^{\prime }\right) =4\pi ^{2}(\delta
(x-x^{\prime })+b_{+}(x,x^{\prime }))
\end{equation*}%
as $x,x^{\prime }\in R_{+}^{4}$ and
\begin{equation*}
-(\Delta -m^{2})G_{-}\left( x,x^{\prime }\right) =4\pi ^{2}b_{-}(x,x^{\prime
})
\end{equation*}%
as $x\in R_{-}^{4}$. The physical region corresponds to $r_{\pm }\geq a$ in
both spaces. Thus we find
\begin{equation*}
G_{\pm }(x,x^{\prime })=G_{0}^{\pm }(x,x^{\prime })+\Delta G_{\pm
}(x,x^{\prime })
\end{equation*}%
where $G_{0}^{-}=0$ since $x^{\prime }\in R_{+}^{4}$ and $G_{0}^{+}=G_{0}$
is the standard Euclidean Green function. The additional sources are
generated on the surface of the sphere $r=a$ and therefore%
\begin{equation*}
b_{\pm }(x,x^{\prime })=\sigma _{\pm }(x,x^{\prime })\frac{1}{a^{3}}\delta
(r-a).
\end{equation*}%
Consider the decomposition%
\begin{equation}
\sigma _{\pm }(x,x^{\prime })=\sum \sigma _{nlm}^{\pm }(a,x^{\prime
})Q_{nlm}(\Omega )  \label{sd}
\end{equation}%
where $\Omega $ is the angular part of the coordinates $x$. Then we find%
\begin{equation*}
\Delta G_{\pm }(x,x^{\prime })=\int G_{0}(x,x^{\prime \prime })\sigma _{\pm
}(a,x^{\prime })\frac{1}{a^{3}}\delta (r^{\prime \prime }-a)d^{4}x^{\prime
\prime }
\end{equation*}%
or substituting the decomposition (\ref{sd}) and
\begin{equation*}
G_{0}(x,x^{\prime })=\sum Q_{nlm}^{\ast }(\Omega ^{\prime })g_{n}^{0}\left(
r,r^{\prime }\right) Q_{nlm}(\Omega )
\end{equation*}%
we find%
\begin{eqnarray*}
\Delta G_{\pm }(x,x^{\prime }) &=&\int \sum Q_{nlm}^{\ast }(\Omega ^{\prime
\prime })g_{n}^{0}\left( r,r^{\prime \prime }\right) Q_{nlm}(\Omega )\times
\\
&&\times \sum^{\prime }\sigma _{n^{\prime }l^{\prime }m^{\prime }}^{\pm
}(a,x^{\prime })Q_{n^{\prime }l^{\prime }m^{\prime }}(\Omega ^{\prime \prime
})\frac{1}{a^{3}}\delta (r^{\prime \prime }-a)d^{4}x^{\prime \prime }
\end{eqnarray*}%
\begin{equation*}
\Delta G_{\pm }(x,x^{\prime })=\sum \sigma _{nlm}^{\pm }(a,x^{\prime
})g_{n}^{0}\left( r,a\right) Q_{nlm}(\Omega )
\end{equation*}%
Comparing this with the solution constructed in the previous section, i.e.,
\begin{equation*}
\Delta G_{\pm }(x,x^{\prime })=\frac{1}{2}\sum \left( \beta _{n}(a,r^{\prime
})\pm \left( 2\alpha _{n}(a,r^{\prime })-\beta _{n}(a,r^{\prime })\right)
\right) Q_{nlm}^{\ast }(\Omega ^{\prime })g_{n}^{0}\left( r,a\right)
Q_{nlm}(\Omega )
\end{equation*}%
we find
\begin{equation*}
\sigma _{nlm}^{\pm }(a,x^{\prime })=\frac{1}{2}\left( \beta _{n}(a,r^{\prime
})\pm \left( 2\alpha _{n}(a,r^{\prime })-\beta _{n}(a,r^{\prime })\right)
\right) Q_{nlm}^{\ast }(\Omega ^{\prime }).
\end{equation*}%
Thus the bias takes the form%
\begin{equation}
b_{\pm }(x,x^{\prime })=\frac{1}{a^{3}}\delta (r-a)\sum_{n=1}^{\infty
}b_{n}^{\pm }(a,r^{\prime })Q_{n}(\widetilde{\chi }),  \label{bias}
\end{equation}%
where
\begin{equation}
b_{n}^{\pm }(a,r^{\prime })=\frac{n}{4\pi ^{2}}\left( \beta _{n}(a,r^{\prime
})\pm \left( 2\alpha _{n}(a,r^{\prime })-\beta _{n}(a,r^{\prime })\right)
\right)  \label{bn}
\end{equation}%
and $\widetilde{\chi }$ is the angle between $x$ and $x^{\prime }$ e.g., see
(\ref{com}). Thus, we see that the true Green function can be obtained by
means of solving the standard equation in the Euclidean space with the
biased source. Indeed, every source generates a set of additional sources on
the inner and outer sides of the throat surface by the rule%
\begin{equation}
\delta (x-x^{\prime })\rightarrow \delta (x-x^{\prime })+b(x,x^{\prime })
\label{b0}
\end{equation}%
where $b(x,x^{\prime })=b_{+}(x,x^{\prime })$ on the outer side of the
throat surface (with respect to the source) and $b(x,x^{\prime
})=b_{-}(x,x^{\prime })$ on the inner side respectively. Since coefficients $%
b_{n}^{\pm }$ include a small parameter $\delta _{n}=\frac{K_{n}(mr^{\prime
})}{mr^{\prime }}/\frac{K_{n}(ma)}{ma}$ which in general case $\delta
_{n}\ll 1$ (e.g., in the massless case it is merely $\left( \frac{a}{%
r^{\prime }}\right) ^{n+1}$), in practical calculations this allows us to
restrict with smallest terms ($n=1,2$) only.

\section{The bias for a wormhole connecting regions in the same space}

Consider now a wormhole which connects two regions in the same Euclidean
space. The rule (\ref{b0}) allows to construct the exact Green function in
this case as well. In this case both spaces $R_{\pm }^{4}$ are glued by the
rule (\ref{gl}) and the true Green function obeys to the standard equation
with the biased source%
\begin{equation*}
-(\Delta -m^{2})G\left( x,x^{\prime }\right) =4\pi ^{2}(\delta (x-x^{\prime
})+b_{tot}(x,x^{\prime }))
\end{equation*}%
where the total bias $b_{tot}$ is constructed by the iteration procedure
\begin{equation}
b_{tot}=\sum_{s=0}^{\infty }b_{s}  \label{btot}
\end{equation}%
as follows. Let $R_{\pm }$ be the positions of the wormhole throat in the
space. Then in the leading order the source $\delta (x-x^{\prime })$
generates the bias in the form%
\begin{equation}
b_{0}(x,x^{\prime })=b_{0}^{+}(x,x^{\prime })+b_{0}^{-}(x,x^{\prime })
\label{ob}
\end{equation}%
where
\begin{equation*}
b_{0}^{\pm }(x,x^{\prime })=\frac{1}{a^{3}}\sum_{n=1}^{\infty }\delta
(r_{\pm }-a)b_{n}^{+}(a,r_{\pm }^{\prime })Q_{n}(\widetilde{\chi }_{\pm
})+\delta (r_{\mp }-a)b_{n}^{-}(a,r_{\pm }^{\prime })Q_{n}(\widetilde{\chi }%
_{\mp }^{\prime })
\end{equation*}%
describes the bias generated with respect to $S_{+}^{3}$ or $S_{-}^{3}$
respectively. We also use here the obvious definitions $r_{\pm }=|x_{\pm
}|=\left\vert x-R_{\pm }\right\vert $, $r_{\pm }^{\prime }=|x_{\pm }^{\prime
}|$, $\cos \widetilde{\chi }_{\pm }=\left( n_{\pm }^{\mu }n_{\pm \mu
}^{\prime }\right) $, and $\cos \widetilde{\chi }_{\pm }^{\prime }=\left(
n_{\pm }^{\nu }(\Lambda ^{\pm 1})_{\nu }^{\mu }n_{\mu \mp }^{\prime }\right)
$ ($n=x/r$). \ This bias takes the structure%
\begin{equation}
b_{0}(x,x^{\prime })=\frac{1}{a^{3}}\rho _{0}^{+}(x,x^{\prime })\delta
(r_{+}-a)+\frac{1}{a^{3}}\rho _{0}^{-}(x,x^{\prime })\delta (r_{-}-a)
\label{r}
\end{equation}%
where $\rho _{0}^{\pm }(x,x^{\prime })=\sum b_{n}^{+}(a,r_{\pm }^{\prime
})Q_{n}(\widetilde{\chi }_{\pm })+b_{n}^{-}(a,r_{\mp }^{\prime })Q_{n}(%
\widetilde{\chi }_{\pm }^{\prime })$ are the surface densities generated on $%
S_{\pm }^{3}$

In the next order every source on $S_{-}^{3}$ generates additional sources
by the same scheme (\ref{ob}) (e.g., see \cite{KS07}) and this defines the
next orders bias as%
\begin{equation*}
b_{s}(x,x^{\prime })=b_{s}^{+}(x,x^{\prime })+b_{s}^{-}(x,x^{\prime })
\end{equation*}%
where%
\begin{equation*}
b_{s}^{\pm }(x,x^{\prime })=\int b_{0}^{\pm }(x,x^{\prime \prime })\rho
_{s-1}^{\mp }(x^{\prime \prime },x^{\prime })\frac{1}{a^{3}}\delta (r_{\mp
}^{\prime \prime }-a)d^{4}x^{\prime \prime }
\end{equation*}%
and $\rho _{s}^{\pm }$ is defined by $b_{s}^{\pm }$ via the relation (\ref{r}%
). \ Thus, we get the iteration scheme for constructing the total bias $%
b_{tot}(x,x^{\prime })$ and the total Green function $G_{total}(x,x^{\prime
})=G_{0}+\int G_{0}(x,x^{\prime \prime })b_{tot}(x^{\prime \prime
},x^{\prime })d^{4}x^{\prime \prime }$. We point out that the decomposition
of the bias (\ref{btot}) includes a small parameter $\delta \leq \frac{%
K_{1}(mR)}{mR}/\frac{K_{1}(ma)}{ma}$ $\leq a^{2}/R^{2}\ll 1$, where $R=|X|$
and $X=R_{+}-R_{-}$ is the distance between throat centers and, therefore,
the iterations converge very rapidly.

\section{The Bias in a gas of wormholes (the rarefied gas approximation)}

The exact form of the bias for a gas of wormholes is constructed by the same
scheme described in the previous section and consists of the two steps.
First, one has to define the bias for a single wormhole (\ref{btot}) and
then one may use iterations. Let $\xi =(a,R_{\pm },\Lambda )$ be wormhole
parameters and, therefore the bias (\ref{btot}) is a function of $\xi $
i.e., $b_{tot}(x,x^{\prime },\xi )$. Then in the first approximation the
bias is merely the sum%
\begin{equation*}
b_{gas}^{0}(x,x^{\prime })=\sum_{A}b_{tot}(x,x^{\prime },\xi ^{A}).
\end{equation*}%
In the next order every first order source generates additional sources
(multiple scattering of the basic signal) as
\begin{equation}
b_{gas}^{1}(x,x^{\prime })=\sum_{A\neq B}\int b_{tot}(x,x^{\prime \prime
},\xi ^{A})b_{tot}(x^{\prime \prime },x^{\prime },\xi ^{B})d^{4}x^{\prime
\prime }  \label{bgas1}
\end{equation}%
etc. By iterating this procedure we may find the exact form of the bias for
the gas of wormholes
\begin{equation}
b_{gas}(x,x^{\prime })=\sum b_{gas}^{s}(x,x^{\prime }).  \label{bgas}
\end{equation}

In the present section we shall consider a dilute gas approximation when
characteristic parameters $a^{2}/r^{\prime 2}$, $a^{2}/R^{2}$, and $%
a^{2}/R_{AB}^{2}\ll 1$. In this case it is sufficient to retain\ in (\ref%
{bias}) the lowest $n=1$ (monopole) term only and neglect mutual scattering
between throats of wormholes (i,e, retain $s=0$ term in (\ref{btot}) and (%
\ref{bgas}), e.g., see for details Ref. \cite{KS07}). Then the bias for a
single wormhole takes the form
\begin{equation}
b^{0}(x,x^{\prime },\xi )=\frac{1}{a^{3}}\delta (r_{+}-a)\left(
b_{1}^{+}(a,r_{+}^{\prime })+b_{1}^{-}(a,r_{-}^{\prime })\right) +\frac{1}{%
a^{3}}\delta (r_{-}-a)\left( b_{1}^{-}(a,r_{+}^{\prime
})+b_{1}^{+}(a,r_{-}^{\prime })\right)   \label{b1}
\end{equation}%
where $b_{1}^{+}$ is given by (\ref{bn}) i.e.,
\begin{equation}
b^{0}(x,x^{\prime },\xi )=\frac{1}{2\pi ^{2}a^{3}}\delta (r_{+}-a)\left(
\alpha _{1}(a,r_{+}^{\prime })-\alpha _{1}(a,r_{-}^{\prime })+\beta
_{1}(a,r_{-}^{\prime })\right) +\frac{1}{2\pi ^{2}a^{3}}\delta
(r_{-}-a)\left( \alpha _{1}(a,r_{-}^{\prime })-\alpha _{1}(a,r_{+}^{\prime
})+\beta _{1}(a,r_{+}^{\prime })\right)
\end{equation}%
where $\alpha _{1}(a,r)$ and $\beta _{1}(a,r)$ are given by (\ref{a}) and (%
\ref{b}).

In the rarefied gas approximation the total bias is additive, i.e.,
\begin{equation}
b_{total}(x,x^{\prime })=\sum b^{0}(x,x^{\prime },\xi _{i})=\int
b^{0}(x,x^{\prime },\xi )F(\xi )d\xi ,  \label{bs}
\end{equation}%
where $F$ is given by%
\begin{equation}
F\left( \xi ,N\right) =\sum\limits_{A=1}^{N}\delta \left( \xi -\xi
_{A}\right) .  \label{F}
\end{equation}%
\begin{equation}
b^{0}(x,x^{\prime })=\int \frac{1}{2\pi ^{2}a^{3}}\delta (\left\vert
x-R_{+}\right\vert -a)\left( \alpha _{1}(a,\left\vert x^{\prime
}-R_{+}\right\vert )-\alpha _{1}(a,\left\vert x^{\prime }-R_{-}\right\vert
)+\beta _{1}(a,\left\vert x^{\prime }-R_{-}\right\vert )\right) F(\xi )d\xi
+(+\leftrightarrow -)
\end{equation}%
Fist, we point out the obvious symmetry of $F$ $(\xi )$ in the replacement $%
(+\leftrightarrow -)$ which means that the two above terms coincide. Then,
we transform $R_{\pm }-x^{\prime }\rightarrow R_{\pm }$ which gives
\begin{equation}
b^{0}(x,x^{\prime })=2\int \frac{1}{2\pi ^{2}a^{3}}\delta (\left\vert
x-x^{\prime }-R_{+}\right\vert -a)\left( \alpha _{1}(a,\left\vert
R_{+}\right\vert )-\alpha _{1}(a,\left\vert R_{-}\right\vert )+\beta
_{1}(a,\left\vert R_{-}\right\vert )\right) F(a,R_{+}+x^{\prime
},R_{-}+x^{\prime })d\xi .
\end{equation}%
For a homogeneous and isotropic distribution $F(\xi )=F(a,X)$, $X=R_{+}-R_{-}
$ then for the bias we find%
\begin{equation}
b^{0}(x-x^{\prime })=2\int \frac{\alpha _{1}(a)}{2\pi ^{2}a^{3}}\delta
(\left\vert x-x^{\prime }-R_{+}\right\vert -a)\left( \frac{K_{1}(mR_{+})}{%
mR_{+}}-\left( 1-\frac{\beta _{1}(a)}{\alpha _{1}(a)}\right) \frac{%
K_{1}(mR_{-})}{mR_{-}}\right) F(a,X)d\xi
\end{equation}%
where $\alpha _{1}(a)\frac{K_{1}(mR_{+})}{mR_{+}}=\alpha _{1}(a,\left\vert
R_{+}\right\vert )$, etc.. For the Fourier transform $b\left( k\right) =\int
b\left( x\right) e^{ikx}d^{4}x$ and $F(a,k)=\int F(a,X)e^{ikx}d^{4}x$ we get
\begin{equation}
b_{total}\left( k\right) =-\frac{4\pi ^{2}}{k^{2}+m^{2}}\nu \left( k\right)
\label{bk}
\end{equation}%
where
\begin{equation}
\nu \left( k\right) =-\int \frac{2\alpha _{1}(a)}{m^{2}}\left( F\left(
a,0\right) -\left( 1-\frac{\beta _{1}(a)}{\alpha _{1}(a)}\right) F\left(
a,k\right) \right) \frac{J_{1}\left( ka\right) }{ka/2}da,  \label{nu(k)}
\end{equation}%
and we used the obvious properties  $m^{2}\frac{K_{1}(mx)}{mx}=\int \frac{%
4\pi ^{2}}{k^{2}+m^{2}}e^{-ikx}\frac{d^{4}k}{\left( 2\pi \right) ^{4}}$ and $%
\int e^{iky}\delta (\left\vert y\right\vert -a)d^{4}y=2\pi ^{2}a^{3}\frac{%
J_{1}\left( ka\right) }{ka/2}$.

The bias (\ref{bk}) is defined in the linear approximation only. Now by the
use of iterations (\ref{bgas1}), (\ref{bgas}) we find $%
N(k)=1+b_{total}(k)+b_{total}^{2}(k)+...$ and  true Green function in a gas
of wormholes becomes
\begin{equation}
G_{true}=G_{0}(k)N(k)=\frac{4\pi ^{2}}{k^{2}+m^{2}+4\pi ^{2}\nu \left(
k\right) }.  \label{tgf}
\end{equation}%
At sufficiently large scales $2\pi /k\gg \ell _{pl}\sim a,\left\vert
R_{+}-R_{-}\right\vert $ we can expand the function $\nu \left( k\right) $
as $\nu \left( k\right) \approx $ $\nu \left( 0\right) +\frac{1}{2}$ $\nu
^{\prime \prime }\left( 0\right) k^{2}$ which defines the renormalized Green
function as%
\begin{equation*}
G_{ren}(k)=\frac{4\pi ^{2}Z}{k^{2}+M^{2}}
\end{equation*}%
where
\begin{equation*}
Z=1/(1+\frac{1}{2}\nu ^{\prime \prime }\left( 0\right) )
\end{equation*}
defines the renormalization of charge values due to the polarization of the
gas of virtual wormholes and
\begin{equation}
M^{2}=(m^{2}+4\pi ^{2}\nu \left( 0\right) )/Z  \label{m}
\end{equation}
that of the mass. We point out that in general $\nu \left( 0\right) $
possesses both signs depending on the sign of $\zeta $. At very small scales
$k\rightarrow \infty $ the function $\nu \left( k\right) $ defines the
spectral density of states of the scalar field \cite{KS10} as $N\left(
k\right) =\left( k^{2}+m^{2}\right) /(k^{2}+m^{2}+4\pi ^{2}\nu \left(
k\right) )$.

The spacetime foam picture assumes that typical virtual wormholes have
throats $a\lesssim \ell _{pl}\sim 10^{-33}cm$ and therefore we can restrict
to the case when $ma\ll 1$. Then in the limit $ma\rightarrow 0$ we find%
\footnote{%
In the opposite case $ma\rightarrow \infty $ (macroscopic virtual wormholes)
we get $\alpha _{1}(a,r^{\prime })=-\left( 1-12\zeta \right) \frac{1}{\sqrt{%
2\pi ma}}e^{ma}$, and $\beta _{1}(a,r^{\prime })=\sqrt{\frac{2ma}{\pi }}%
e^{ma}$.} $\ \alpha _{1}(a)=-\frac{1-6\zeta }{2\left( 1-3\zeta \right) }%
m^{2}a^{2}$, $\beta _{1}(a)=\frac{6\zeta }{2\left( 1-3\zeta \right) }%
m^{2}a^{2}$ and the function $\nu \left( k\right) $ becomes
\begin{equation}
\nu \left( k\right) =-\int a^{2}\frac{F\left( a,k\right) -\left( 1-6\zeta
\right) F\left( a,0\right) }{\left( 1-3\zeta \right) }\frac{J_{1}\left(
ka\right) }{ka/2}da.  \label{nu}
\end{equation}%
The non-minimal coupling $\zeta \neq 0$ generates an additional source at
wormhole throats, which in the case $m=0$ brakes the Gauss theorem ($\oint
\frac{\partial G}{\partial n}dS=-4\pi ^{2}$). Therefore, it is convenient to
split the bias into two parts $\nu \left( k,\zeta \right) =\nu \left(
k,0\right) +\mu \left( k,\zeta \right) $ where the first part is
\begin{equation}
\nu \left( k,0\right) =\int a^{2}\left( F\left( a,0\right) -F\left(
a,k\right) \right) \frac{J_{1}\left( ka\right) }{ka/2}da
\end{equation}%
and the second part is
\begin{equation}
\mu \left( k,\zeta \right) =-\frac{3\zeta }{\left( 1-3\zeta \right) }\int
a^{2}\left[ \left( F\left( a,k\right) +F\left( a,0\right) \right) \right]
\frac{J_{1}\left( ka\right) }{ka/2}da.
\end{equation}%
In the long wave limit $k\rightarrow 0$ one gets $F\left( a,k\right) \approx
F\left( a,0\right) +\frac{1}{2}F^{\prime \prime }\left( a,0\right) k^{2}$
(for the isotropic distribution of wormholes $F^{\prime }\left( a,0\right) =0
$ and $F^{\prime \prime }\left( a,0\right) <0$) and one gets
\begin{equation}
\nu \left( k,0\right) \approx k^{2}\int a^{2}\frac{\left( -F^{\prime \prime
}\left( a,0\right) \right) }{2}da
\end{equation}%
i.e., $\nu \left( k,0\right) \rightarrow 0$, as $k\rightarrow 0$ which in
the massless ($\zeta =0$) case defines a partial screening of the source
(e.g., see \cite{KS10}) $N\left( 0\right) \rightarrow 1/(1+2\pi ^{2}\int
a^{2}\left( -F^{\prime \prime }\left( a,0\right) \right) da)$. The second
part may describe the so-called dark source
\begin{equation*}
\mu \left( 0,\zeta \right) =-\frac{6\zeta }{\left( 1-3\zeta \right) }\int
a^{2}F\left( a,0\right) da
\end{equation*}%
whose sign depends on the value of $\zeta $ and which generates or merely
renormalizes the mass (\ref{m}).

\section{Summary}

Consider first \textit{\ an example of a finite density of wormholes.}

In the case of a finite density of wormholes the space remains to be
Euclidean at very small scales and therefore $N(k)\rightarrow 1$ as $%
k\rightarrow \infty $. To illustrate this we consider now a particular form
for $F\left( a,X\right) $, e.g.,
\begin{equation}
F(a,X)=\frac{n}{2\pi ^{2}r_{0}^{3}}\delta (a-a_{0})\delta (X-r_{0}),
\label{ff}
\end{equation}%
where $n=N/V$ is the density of wormholes. This function corresponds to a
set of wormholes with the fixed throat size $a_{0}$ and the distance between
throats $r_{0}=\left\vert R_{+}-R_{-}\right\vert $. Then $F(a,k)=\int
F(a,X)e^{ikx}d^{4}x$ reduces to $F(a,k)=n\frac{J_{1}(kr_{0})}{kr_{0}/2}%
\delta (a-a_{0})$. Thus from (\ref{bk}) we find
\begin{equation}
\nu \left( k\right) =-2n\frac{\alpha _{1}(a_{0})}{m^{2}}\left( 1-\left( 1-%
\frac{\beta _{1}(a_{0})}{\alpha _{1}(a_{0})}\right) \frac{J_{1}(kr_{0})}{%
kr_{0}/2}\right) \frac{J_{1}\left( ka_{0}\right) }{ka_{0}/2}.
\end{equation}%
In the limit $ma\ll 1$ we get (compare with the analogous expression in \cite%
{KS10})
\begin{equation}
\nu \left( k,0\right) =na_{0}^{2}\left( 1-\frac{J_{1}(kr_{0})}{kr_{0}/2}%
\right) \frac{J_{1}\left( ka_{0}\right) }{ka_{0}/2}
\end{equation}%
\begin{equation}
\mu \left( k,\zeta \right) =-\frac{3\zeta }{\left( 1-3\zeta \right) }%
na_{0}^{2}\left( 1+\frac{J_{1}(kr_{0})}{kr_{0}/2}\right) \frac{J_{1}\left(
ka_{0}\right) }{ka_{0}/2}
\end{equation}%
As $k\rightarrow 0$ we get $J_{1}(kr_{0})/\frac{kr_{0}}{2}\approx 1-\frac{1}{%
2}(\frac{kr_{0}}{2})^{2}+...$ which gives
\begin{equation*}
\nu (k)\approx \frac{na^{2}r_{0}^{2}}{8\left( 1-3\zeta \right) }k^{2}-\frac{%
6\zeta na^{2}}{1-3\zeta }.
\end{equation*}%
Which defines $Z=1/\left( 1+\frac{na^{2}r_{0}^{2}}{8\left( 1-3\zeta \right) }%
)\right) $ and $M^{2}=(m^{2}-4\pi ^{2}na^{2}\frac{6\zeta }{1-3\zeta })/Z$ .
Thus, we see that in the long-wave limit ($ka$, $kr_{0}\ll 1$) the presence
of a particular set of virtual wormholes renormalizes merely the value of
the charge and mass values. In the short-wave limit $k\rightarrow \infty $
we get $\frac{J_{1}\left( z\right) }{z/2}\approx \sqrt{\frac{8}{\pi z^{3}}}%
\cos \left( z-3\pi /4\right) +...$ and $\nu \left( k\right) \rightarrow 0$
as
\begin{equation}
\nu \left( k\right) \approx \frac{na^{2}}{\left( 1-3\zeta \right) }\left(
1-6\zeta -\sqrt{\frac{8}{\pi \left( kr_{0}\right) ^{3}}}\cos \left( kr_{0}-%
\frac{3\pi }{4}\right) \right) \sqrt{\frac{8}{\pi \left( ka_{0}\right) ^{3}}}%
\cos \left( ka_{0}-\frac{3\pi }{4}\right) ,
\end{equation}%
i.e., the bare parameters of the field theory restore $Z\rightarrow 1$ and $%
M\rightarrow m$ which means that in the presence of a finite density of
wormholes at sufficiently small scales space looks like the ordinary
Euclidean space.

\textit{The case of limiting topologies or infinite densities of wormholes}

Consider now the limiting distribution when the density of wormholes $%
n\rightarrow \infty $. Since every throat cuts the finite portion of the
volume $\frac{\pi ^{2}}{2}a^{4}$, this case requires considering the limit $%
a\rightarrow 0$, i.e., wormhole throats degenerate into points and the
minimal scale below which the space looks like the Euclidean space is merely
absent. We assume that in this limit $a^{2}F(a,X)=\delta \left( a\right)
f\left( X\right) $ where $f\left( X\right) $ is a finite specific
distribution. Then (\ref{nu}) reduces to
\begin{equation*}
\nu \left( k\right) =\left( \widetilde{f}\left( 0\right) -\widetilde{f}%
\left( k\right) \right) -\frac{3\zeta }{\left( 1-3\zeta \right) }\left(
\widetilde{f}\left( k\right) +\widetilde{f}\left( 0\right) \right)
\end{equation*}%
where $\widetilde{f}\left( k\right) =\int f\left( X\right) e^{ikX}d^{4}X$
(in particular, for conformal coupling $\zeta =\frac{1}{6}$ we get simply $%
\nu \left( k\right) =-2\widetilde{f}\left( k\right) $). Now if we require
that $\widetilde{f}\left( k\right) $ is chosen in such a way that
\begin{equation*}
G_{true}\left( x-x^{\prime }=0\right) =\int \frac{4\pi ^{2}}{%
k^{2}+m^{2}+4\pi ^{2}\nu \left( k\right) }\frac{d^{4}k}{\left( 2\pi \right)
^{4}}<\infty ,
\end{equation*}%
then quantum field theory in such spaces turns out to be finite.
We recall that upon continuation to the Minkowsky space the
Euclidean Green function transforms to the Feynman propagator
$G(x-x^{\prime })\rightarrow 4\pi ^{2}D_{F}(x-x^{\prime })$. In
other words we get here a class of limiting topologies where Green
functions (\ref{tgf}) $G_{true}$ or the Feynman propagator $D_{F}$
have a good ultraviolet behavior and QFT is free of divergencies,
e.g., see \cite{L,book}. We also point out that the explicit form
of $\nu \left( k\right) $ can be found numerically in the lattice
quantum gravity e.g., see \cite{AJL,MRS,LQG} and references
therein.

\section{Acknowledgment}

I would like to acknowledge A. Kirillov for valuable discussions and
comments.

\end{document}